\documentclass[twocolumn]{aastex631}

\shorttitle{Barred versus Unbarred Narrow-Line Seyfert 1}
\shortauthors{Yu}
\graphicspath{{./}{figures/}}
\usepackage{subfigure}
\usepackage{xurl}
\begin{document}

\title{Large-Scale Stellar Bars and Nuclear Activity in Narrow-Line Seyfert 1 Galaxies}

\author[0000-0001-8894-0854]{Po-Chieh Yu}
\affiliation{Taiwan Astronomical Research Alliance (TARA), Taipei, Taiwan}
\altaffiliation{https://tara.tw}
\affiliation{Institute of Astronomy and Astrophysics, Academia Sinica, No. 1, Sec. 4, Roosevelt Road, 106319, Taipei, Taiwan}
\email{pcyu@asiaa.sinica.edu.tw}  

\begin{abstract}
We present a statistical comparison between barred and unbarred narrow-line Seyfert 1 galaxies (NLS1s) to investigate the role of large-scale stellar bars
in fueling AGN with low-mass black holes. From a parent sample of 492 NLS1s selected from SDSS and Pan-STARRS1 (PS1), we construct a PS1-resolved matched
sample of 101 barred and 101 unbarred NLS1s, matched in redshift, inclination, and image quality. Unbarred NLS1s are modestly but consistently brighter than
barred NLS1s in several optical AGN luminosity indicators, including $\lambda L_{\lambda}(5100~\mbox{\AA})$, broad H$\alpha$, broad H$\beta$,
narrow H$\alpha$, narrow H$\beta$, and [\ion{O}{3}]~$\lambda 5007$, with typical offsets of $\sim 0.1$--$0.25$ dex; the broad H$\alpha$ and narrow-line
offsets are statistically significant, while the continuum and broad H$\beta$ offsets are marginal. Monte Carlo resampling and a Bayesian model support the
same direction of offset. In contrast, the main narrow-line diagnostics, including [\ion{O}{3}]/H$\beta$, O3N2, and the [\ion{S}{2}]-based electron
density, are statistically consistent between the subsamples, although barred NLS1s show a higher N2 ratio. Within the barred subsample, the bar strength
parameter $e_{\max}$ shows a mild anti-correlation with the optical continuum and broad Balmer-line luminosities. These results indicate that large-scale stellar bars are not associated with enhanced instantaneous optical AGN luminosity. Instead, barred systems may experience more regulated or time-variable inner gas delivery, building circumnuclear gas reservoirs without necessarily increasing accretion power observed at the epoch traced by optical AGN emission.
\end{abstract}

\keywords{Galaxy bars (2364) --- Seyfert galaxies (1447) --- Active galactic nuclei (16) --- Galaxy structure (622) --- Galaxy evolution (594)}

\section{Introduction} \label{sec:intro}
Active galactic nuclei (AGNs) may play a central role in the coevolution of galaxies and their central supermassive black holes \citep{Kormendy2013,Heckman2014}. A major open question in AGN research is how material is transported from galactic scales into the nuclear regions, where it can fuel accretion onto the central black hole \citep{Combes2019,Audibert2019}. Two primary mechanisms have been proposed: large-scale stellar bars \citep{Shlosman1989,Ho1997} and galaxy mergers \citep{Sanders1988,Barnes1991}. In particular, stellar bars are thought to efficiently redistribute angular momentum and funnel gas inward from kiloparsec scales to the central few hundred parsecs, where it may trigger both nuclear star formation and AGN activity \citep{Ho1997,Martin1997,Sakamoto1999,Hopkins2010,Ellison2011}.

The influence of large-scale bars on AGNs, however, remains controversial. Some studies have found that bar fractions are similar in AGN hosts and inactive galaxies \citep{Ho1997,Hunt1999,Cisternas2015,Zee2023}, whereas others reported that AGNs show enhanced bar fractions \citep{Knapen2000,Laurikainen2004,SilvaLima2022}. From the perspective of AGN strength, \citet{Alonso2013,Alonso2014,Alonso2018} and \citet{Marels2025} showed that barred AGNs tend to have higher [O III] luminosities and younger stellar populations than their unbarred counterparts. In contrast, \citet{Cisternas2013,Cisternas2015} and \citet{Galloway2015} reported no significant difference in nuclear activities, and \citet{Goulding2017} concluded that bars do not significantly affect black hole accretion rates in nearby galaxies ($z$ $<$ 0.15). 

Beyond the presence of a bar, its strength may also play an important role. Based on a $^{12}\mathrm{CO}(J=1$--$0)$ mapping survey, \citet{Kuno2007} found higher central molecular gas concentrations in barred galaxies and showed that the central gas concentration is related to bar properties. \citet{Yu2022} further found that stronger bars are more efficient at transporting cold gas toward the central regions of galaxies.

While some studies have linked strong bars to enhanced star formation \citep{Aguerri1999}, others have associated them with reduced star formation efficiency, older stellar populations, redder colors, and higher metallicities \citep{Vera2016,Zee2023}. These results suggest that both the presence and strength of bars may influence AGN fueling and host galaxy evolution, though the overall picture remains an area of ongoing research.

AGNs are believed to be powered by accretion disks surrounding supermassive black holes, with different types exhibiting distinct emission-line properties \citep{Khachikian1974,Osterbrock1981}. Narrow-line Seyfert 1 galaxies (NLS1s) are a unique class of AGNs \citep{Osterbrock1985}. Unlike typical Seyfert 1 galaxies (broad-line Seyfert 1; BLS1s), which display broad H$\beta$ emission lines with FWHM greater than 2000 km/s, NLS1s are characterized by relatively narrower broad permitted emission lines with FWHM less than 2000 km/s \citep{Goodrich1989}. They stand out due to their lower supermassive black hole (SMBH) masses, typically ranging from 10$^{6}$ -- 10$^{8}$$M_{\odot}$, which is significantly less than those found in other Seyfert galaxies and many quasars \citep{Peterson2000,Zhou2006}. These galaxies also exhibit high Eddington ratios, indicating efficient and rapid black hole growth \citep{Mathur2000}. The combination of large-scale environments, lower black hole masses and high Eddington ratios suggests that NLS1s may represent an early phase in AGN evolution, where black holes are actively accreting matter before evolving into more typical Seyfert galaxies or quasars \citep{Jarvela2017,Kim2019,Mathur2000,Winkel2022}. 

Previous studies have predominantly examined Seyfert 1, Seyfert 2 galaxies, and LINERs, which typically feature high black hole masses and lower accretion rates, indicating they are in more advanced evolutionary stages. However, the influence of large-scale bars on AGNs may not be prominently observed or strongly impactful at these stages. Focusing on large-scale bars of NLS1s offers a promising way to deepen our understanding of how AGNs form and evolve, potentially shedding light on the role of large-scale bars in this process. 

Several studies have indeed suggested a strong connection between bars and NLS1s. \citet{Kim2017} reported that the bar fraction among NLS1s (48\%) is significantly higher than that among BLS1s (22\%), consistent with earlier results \citep{Crenshaw2003,Deo2006,Ohta2007,deXivry2011}. However, these works were limited by small sample sizes, often fewer than 50 NLS1s, leaving the statistical significance uncertain. To establish a clearer picture, larger and more systematic studies are required, combining reliable bar classifications with multi-wavelength diagnostics of AGN activity.

Motivated by (1) the possibility that NLS1s represent an early phase of AGN evolution, (2) the scarcity of large-sample studies linking host galaxy morphology with the properties of NLS1s, and (3) the need to investigate the role of bars in AGNs at early stages, this work presents a statistical comparison between barred and unbarred NLS1 galaxies. Using Pan-STARRS \citep[PS1;][]{Chambers2016,Flewelling2020,Magnier2020a,Magnier2020b,Magnier2020c,Waters2020} imaging, we classify bar structures in NLS1 hosts, while Sloan Digital Sky Survey \citep[SDSS;][]{York2000} spectra provide optical emission-line diagnostics. By combining host morphology and spectroscopic measurements, this study aims to clarify the role of bars in fueling AGNs and shaping the evolutionary pathways of NLS1 galaxies.

The paper is organized as follows. In Section 2, we describe the sample selection, bar identification, and matched subsample pairing. Section 3 presents the statistical analysis of optical spectra for barred and unbarred NLS1s. Section 4 discusses the implications of our results for AGN fueling. Finally, Section 5 summarizes our main conclusions.

\section{Data and Method} 
\subsection{Imaging Data}

The PS1 all-sky survey provides extensive imaging datasets in the $g$, $r$, $i$, $z$, and $y$ bands for morphological studies
\citep{Chambers2016,Flewelling2020}. Among these, $g$-band images are more affected by dust extinction, while the stacked $y$-band data are relatively
shallow, with a mean $5\sigma$ point-source limiting sensitivity of $\sim 21.4$ mag \citep{Chambers2016}. For this work, we therefore employ
PS1 $r$-band stacked images for the morphological classification of NLS1 host galaxies. In the matched sample, these data have a typical effective
PSF size of $\sim 1\farcs1$ (Table~1), compared with the median $r$-band seeing of $\sim 1\farcs4$ in SDSS imaging \citep{Lupton2001}. The PS1 stacked $r$-band data also reach a mean $5\sigma$ point-source limiting sensitivity of $\sim 23.2$ mag \citep{Chambers2016}. This combination provides uniform photometric coverage for the SDSS spectroscopic NLS1 sample.

A comprehensive catalog of NLS1s from the SDSS Data Release 17 \citep[DR17;][]{DR172022} was presented by \citet{Paliya2024}, containing 22,656 objects. 
This sample is nearly two times larger than previous NLS1 compilations \citep{Rakshit2017}. 
Instead of using the H$\beta$ FWHM threshold of 2200 km/s without considering the uncertainties \citep{Rakshit2017}, \citet{Paliya2024} followed the traditional definition of NLS1 galaxies, e.g., the FWHM of the broad H$\beta$ emission line within measured uncertainties is smaller than 2000 km/s \citep{Osterbrock1985}.
The catalog provides key AGN spectral properties, including nuclear monochromatic luminosity at 5100 \AA, luminosities of broad and narrow lines, velocity information, and accretion properties. This dataset is therefore well suited to extend the NLS1 sample for the present study.

We cross-matched the catalog of \citet{Paliya2024} with the PS1 survey and retrieved $r$-band stacked image cutouts through the STScI PS1 Image Cutout Service\footnote{\url{https://ps1images.stsci.edu/cgi-bin/ps1cutouts}}. At a representative redshift of $z = 0.1$, this corresponds to a physical resolution of $\sim$1.8 kpc/arcsec, which is sufficient to resolve typical large-scale bars with sizes of 3–5 kpc. In total, 492 NLS1 galaxies with redshifts $< 0.1$ were identified within the PS1 footprint. These objects constitute the parent sample for our bar identification analysis.

\subsection{Bar Identification}
\begin{figure*}[ht!]
  \centering
  \includegraphics[width=\textwidth]{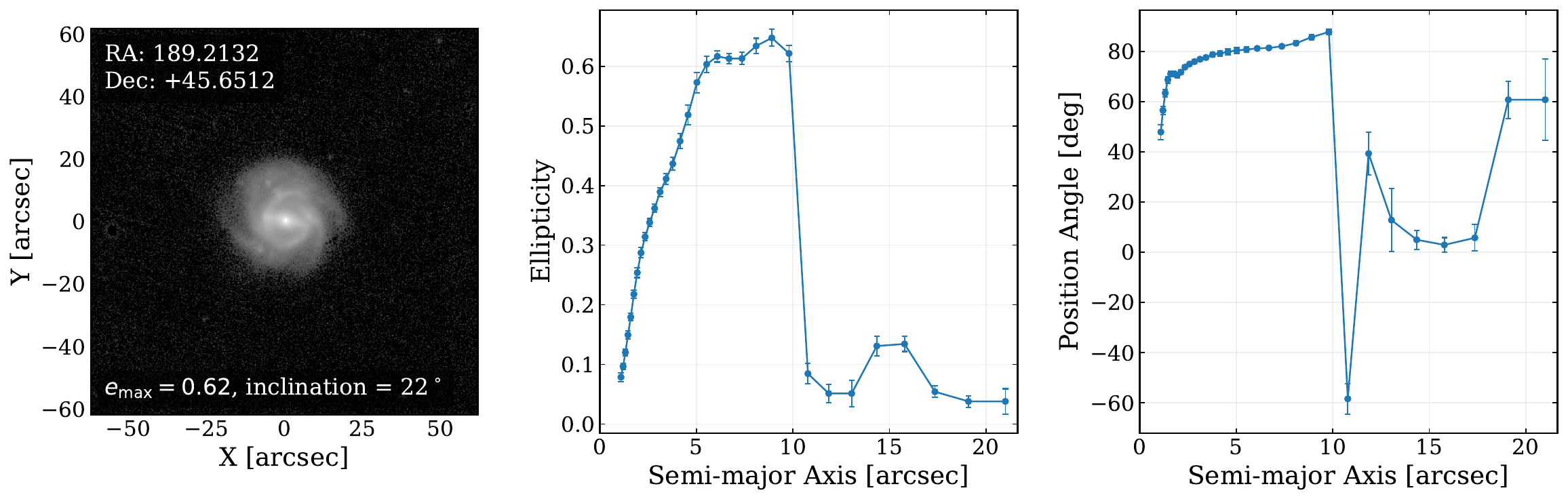}
\caption{An example of the bar-identification procedure. Left: PS1 $r$-band galaxy image. Middle: ellipticity ($e$) as a function of semi-major axis (SMA). Right: position angle (PA) as a function of SMA. Here, $e_{\max}$ denotes the first qualifying local maximum in the ellipticity profile adopted by our conservative bar-identification procedure, reducing potential contamination from structures beyond the bar.}
  \label{fig:fig1}
\end{figure*}

We identified bar structures using the \textsc{ellipse} task \citep{Jedrzejewski1987} in \textsc{IRAF} \citep{Tody1986,Tody1993}, which has been widely employed to characterize bars by measuring ellipticity ($e$), position angle (PA), and surface brightness as a function of radius. Bars are characterized by isophotes that remain elongated with nearly constant orientation across the bar region, followed by a marked decrease in ellipticity and a significant change in PA at the bar end. Accordingly, we adopted the bar identification criteria of \citet{Marinova2007}: (a) the maximum ellipticity ($e_{\max}$) of a bar must exceed 0.25; (b) the PA should remain stable within 10$^\circ$ across the bar region; and (c) at the bar end, the ellipticity should decrease by $>0.1$ accompanied by a PA change of $>10^\circ$. We masked bright stars on the images and then fitted ellipse to each image out to where the galaxy isophotes reach the sky level. An example is shown in Figure~\ref{fig:fig1}.

Bar structures could be misidentified due to projection effects of high inclination angles.
We estimated the inclination angles using the disk axis ratio, adopting the standard approximation:
$\cos^{2} i = \frac{(b/a)^{2} - q_{0}^{2}}{1 - q_{0}^{2}}$,  $b/a$ is the disk axis ratio, $q_{0}$ = 0.2 is the assumed intrinsic thickness of spiral galaxies \citep{Holmberg1958,Padilla2008}. The disk axis ratio $(b/a)$ was determined from the ellipticity measured in the outer isophotes 
semi-major axes SMA $\geq 0.6 \, $SMA$_{\mathrm{max}}$, where SMA$_{\mathrm{max}}$ is the largest fitted semi-major axis. 
Galaxies with inclination angles $i>60^\circ$ were excluded, as bar identification is more reliable in moderately inclined disk galaxies \citep[e.g.,][]{Marinova2007,Sheth_2008}.

To ensure that bar classifications were based on sufficiently resolved galaxy structures, we required SMA$_{\max}$ to be larger than four times the effective PS1 resolution, $\mathrm{FWHM}_{\mathrm{eff}}$:

\[
\frac{\mathrm{SMA}_{\max}}{\mathrm{FWHM}_{\mathrm{eff}}} > 4,
\]

where the $\mathrm{FWHM}_{\mathrm{eff}}$ for each image is the geometric mean of the major and minor FWHM values in the PS1 \texttt{StackObjectView} table \citep{Flewelling2020}:
\[
\mathrm{FWHM}_{\mathrm{eff}} = \sqrt{\mathrm{FWHM}_{\mathrm{major}} \times \mathrm{FWHM}_{\mathrm{minor}}}.
\]
This provides a single representative value of the PSF size and accounts for anisotropic seeing.
Galaxies that did not satisfy this criterion were classified as unresolved and were excluded from the matched barred--unbarred comparison.

This criterion is intentionally conservative. Previous studies suggest that stellar bars can generally be identified when their characteristic length exceeds roughly $\sim2$--$3$ times the PSF FWHM \citep[e.g.,][]{Aguerri2009,Liang2024}. However, our quantity SMA$_{\max}$ is not a direct measurement of the bar length; rather, it traces the maximum radial extent over which reliable isophotal fitting can be performed. We therefore use SMA$_{\max}$/$\mathrm{FWHM}_{\mathrm{eff}} > 4$ as a conservative resolution requirement for identifying bar structure, rather than as a direct translation of the literature bar-length threshold. It should be noted that intrinsically short bars may still be missed because of PSF smearing even when the galaxy satisfies SMA$_{\max}$/$\mathrm{FWHM}_{\mathrm{eff}}>4$. This effect becomes more
important at higher redshift, where bars of a given physical size have smaller angular sizes. In particular, bars with radii smaller than approximately twice the FWHM become difficult to identify \citep{Erwin2018,Liang2024}.

\begin{figure}[ht!]
  \centering
  \includegraphics[width=\columnwidth]{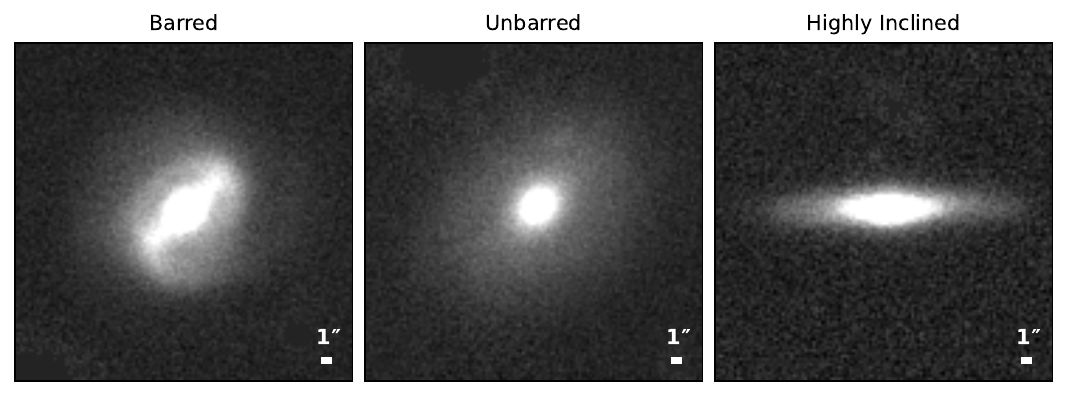}
  \caption{Pan-STARRS cutouts of representative examples of barred, unbarred, and inclined systems.}
  \label{fig:fig2}
\end{figure}

Following the \textsc{ellipse} fitting results, our sample consists of 105 barred, 208 unbarred, 42 highly inclined, and 137 unresolved NLS1s. Representative examples of barred, unbarred, and inclined systems are shown in Figure~\ref{fig:fig2}.

\subsection{Subsample}

The barred and unbarred samples exhibit significantly different redshift distributions, reflecting the well-known selection bias against bar detection at higher redshifts. To mitigate this bias, as well as the effects of galaxy inclination, we construct a matched subsample by pairing each barred galaxy with the nearest unbarred counterpart in the two-dimensional redshift–inclination parameter space using the \textit{NearestNeighbors} algorithm implemented in \textit{scikit-learn} \citep{pedregosa2011}. Each barred galaxy is paired with the nearest unbarred galaxy without replacement, resulting in 101 matched pairs (Table~\ref{tab:barred_sample} and Table~\ref{tab:unbarred_sample}). Because the parent sample is drawn from a flux-limited survey, matching in redshift ensures that both subsamples are subject to the same luminosity selection function.

\begin{figure}[ht!]
  \centering
  \includegraphics[width=\columnwidth]{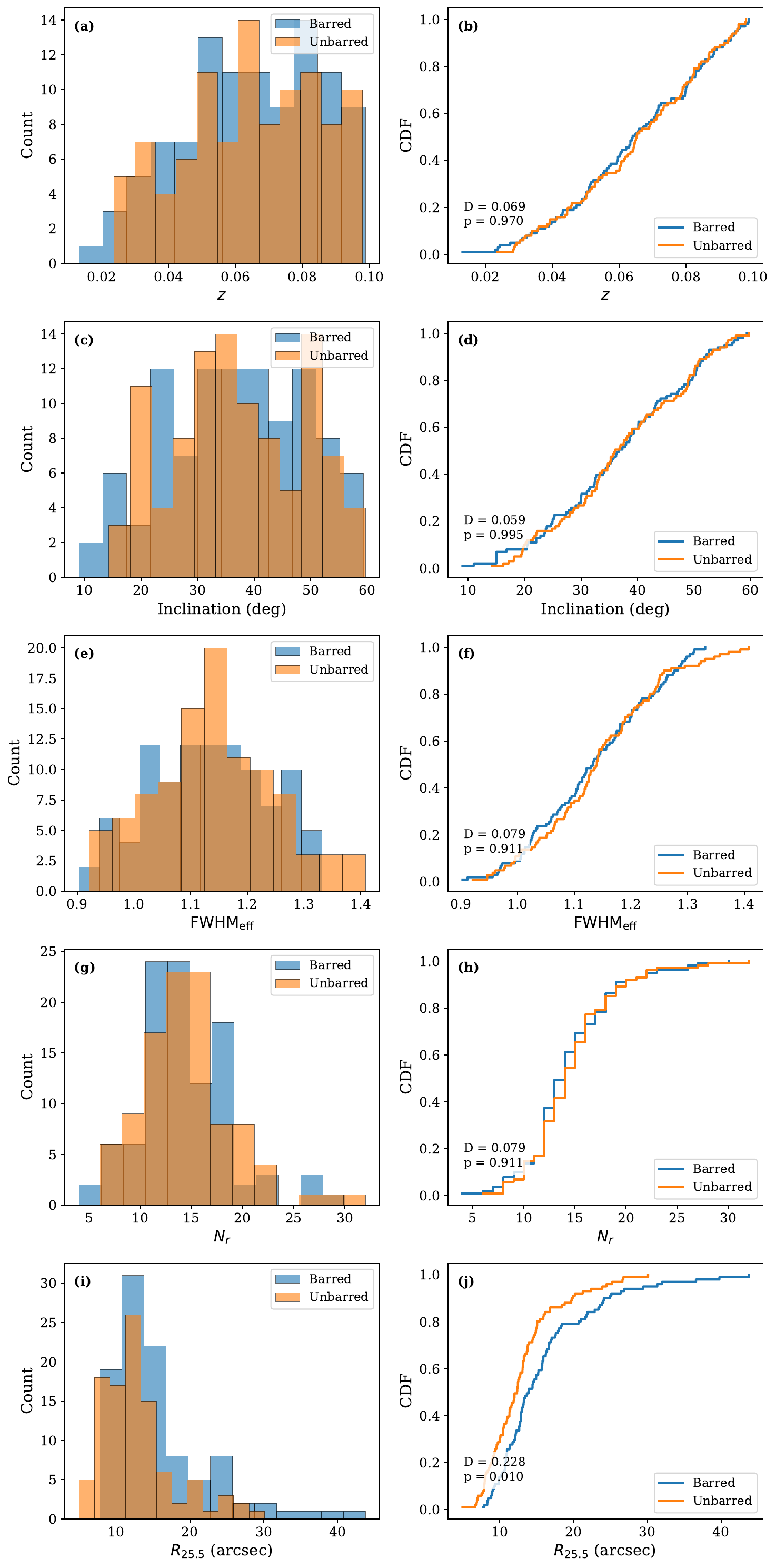}
  \caption{Comparison of the basic properties between the barred and unbarred NLS1 subsamples. The left panels show histograms and the right panels show the corresponding cumulative distribution functions (CDFs). Panels (a) and (b) present the redshift distribution, panels (c) and (d) show the inclination angle, panels (e) and (f) show the effective PSF size $\mathrm{FWHM}_{\mathrm{eff}}$, panels (g) and (h) show the number of stacked $r$-band images $N_r$, and panels (i) and (j) present the $R_{25.5}$ distribution.} The Kolmogorov–Smirnov test statistics are indicated in the CDF panels.
 \label{fig:fig3}
\end{figure}

A Kolmogorov–Smirnov (K–S) test confirms that the redshift and inclination distributions of the two subsamples are statistically indistinguishable ($p \approx 0.99$). We further compare the distributions of the number of stacked $r$-band exposures ($N_{r}$) and the $\mathrm{FWHM}_{\mathrm{eff}}$ to ensure comparable flux depth and spatial resolution. Both K–S and Welch $t$-test indicate no significant differences between the two subsamples (Figure~\ref{fig:fig3}).

We also compare the angular sizes of the two subsamples using $R_{25.5}$, defined as the semi-major axis at which the $r$-band surface
brightness reaches 25.5 mag arcsec$^{-2}$, as measured from the IRAF \textit{ellipse} profiles. As shown in Figure~\ref{fig:fig3}, the barred
galaxies have larger angular sizes than the unbarred galaxies, with median $R_{25.5}$ values of 13\farcs90 and 12\farcs35, respectively. 
The difference is statistically significant (K--S $p=0.010$). Thus, although the two subsamples have comparable $\mathrm{FWHM}_{\mathrm{eff}}$ distributions, their relative spatial resolution is not identical. In summary, the barred and unbarred subsamples have consistent redshifts, inclinations, and image quality, while the barred galaxies have systematically larger angular sizes and hence somewhat better relative spatial resolution. Because the fixed 3-arcsec SDSS fiber therefore samples different relative fractions of the host galaxies, aperture effects could potentially influence the measured narrow-line emission \citep{Kewley2005,Maragkoudakis2014}. The possible impact of this difference on the narrow-line comparison is examined in Section~\ref{sec:Dis}.

\begin{figure}[ht!]
  \centering
  \includegraphics[width=\columnwidth]{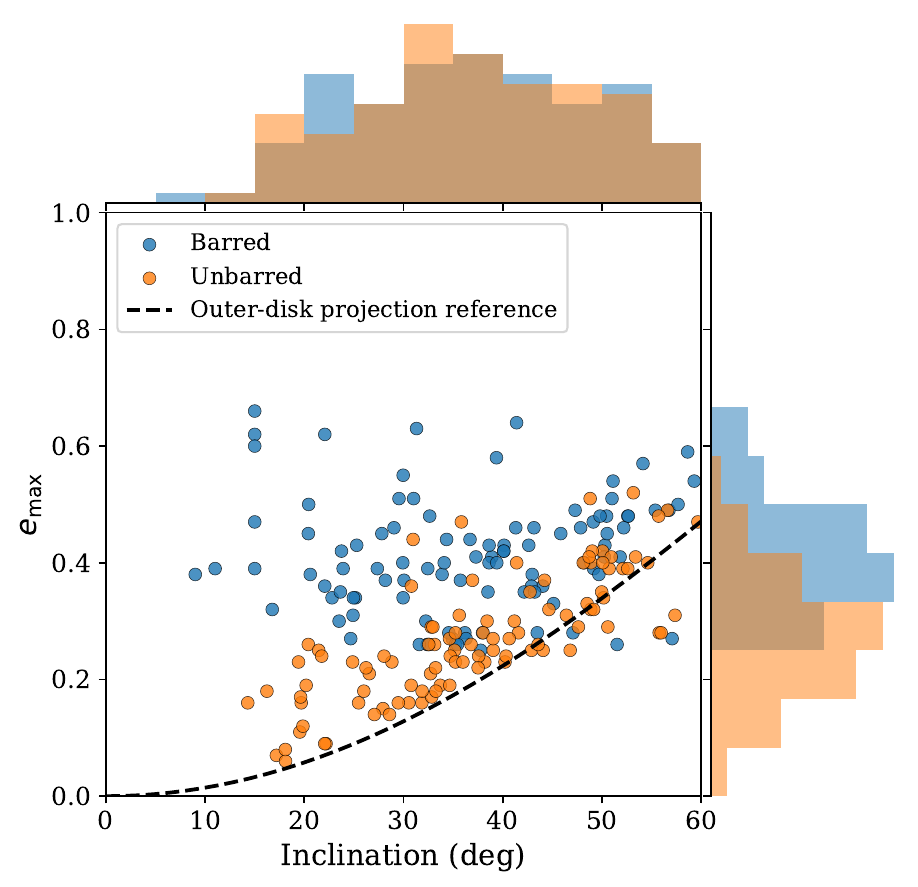}
  \caption{The maximum ellipticity as a function of galaxy inclination for the barred and unbarred NLS1 subsamples. The histograms show the distributions of inclination and ellipticity. The ellipticity peak in unbarred galaxies closely follows the expected disk projection relation, whereas barred galaxies show significantly higher ellipticities and a weaker dependence on inclination.}
\label{fig:fig4}
\end{figure}

To investigate potential projection effects in the bar identification, we further examine the relation between maximum ellipticity and galaxy inclination. Since inclinations are derived from the ellipticity of the outer disk, the $e_{\max}$ values of unbarred galaxies are expected to follow the apparent ellipticity of a projected disk. The dashed curve in Figure~\ref{fig:fig4} corresponds to the expected relation as mentioned in subsection~2.2. The unbarred galaxies approximately follow this trend, indicating that their measured ellipticity is dominated by disk inclination. In contrast, barred galaxies show a much weaker dependence on inclination and typically exhibit larger ellipticities than expected from projection alone, suggesting that the ellipticity peak primarily traces the bar structure rather than the projected disk.

\section{Spectroscopic Comparison and Statistical Analysis}

We compared catalog-provided spectroscopic data from \citet{Paliya2024} for barred and unbarred NLS1s with K-S test and Welch $t$-test. All statistical results are listed in Table~\ref{tab:tab1}. Significance tests were further verified with Monte Carlo resampling and Bayesian hierarchical modeling \citep{Salvatier2016}. 

The Monte Carlo resampling analysis explicitly accounts for measurement uncertainties. For each galaxy, the observed luminosity was perturbed by drawing a random value from a Gaussian distribution centered on the measured value with a standard deviation equal to its quoted uncertainty, assuming that the measurement errors are independent and well described by normal distributions. This procedure propagates measurement uncertainties but does not account for potential systematic errors or sample selection effects. For each realization, we recomputed the mean luminosity difference between the barred and unbarred subsamples, as well as the corresponding Welch $t$-test. This procedure was repeated 10,000 times for each luminosity indicator.

We also modeled the luminosity differences with a simple Bayesian hierarchical model. For each luminosity quantity, the observed value $y_i$
was treated as a noisy measurement of a latent true luminosity $\theta_i$ with observational uncertainty $\sigma_i$. The latent luminosities were
drawn from barred and unbarred parent distributions with group means $\mu_{\rm bar}$ and $\mu_{\rm unbar}$ and a common intrinsic scatter $\tau$.
We then inferred the posterior distribution of $\Delta=\mu_{\rm unbar}-\mu_{\rm bar}$ and report the posterior mean of $\Delta$ and $P(\Delta>0)$. 
The model was implemented with \texttt{PyMC} \citep{Salvatier2016}, using the No-U-Turn Sampler \citep[NUTS;][]{Hoffman2014}. 
We ran four independent chains with 2000 tuning steps and 10,000 posterior draws per chain. This analysis accounts for individual measurement uncertainties and intrinsic population scatter, and provides a probabilistic consistency check on the Monte Carlo results.

\begin{deluxetable*}{lccccccccccc}
\tablecaption{Properties of the Matched Barred NLS1 Sample
\label{tab:barred_sample}}
\tablehead{
\colhead{Pair} &
\colhead{SDSS Name} &
\colhead{R.A.} &
\colhead{Decl.} &
\colhead{$z$} &
\colhead{$i$} &
\colhead{$\mathrm{FWHM}_{\mathrm{eff}}$} &
\colhead{$N_r$} &
\colhead{$R_{25.5}$} &
\colhead{$e_{\max}$} &
\colhead{$a_{\rm bar}$} &
\colhead{$a_{\rm bar}/\mathrm{FWHM}_{\mathrm{eff}}$} \\
\colhead{} &
\colhead{} &
\colhead{(deg)} &
\colhead{(deg)} &
\colhead{} &
\colhead{(deg)} &
\colhead{(arcsec)} &
\colhead{} &
\colhead{(arcsec)} &
\colhead{} &
\colhead{(arcsec)} &
\colhead{} \\
\colhead{(1)} &
\colhead{(2)} &
\colhead{(3)} &
\colhead{(4)} &
\colhead{(5)} &
\colhead{(6)} &
\colhead{(7)} &
\colhead{(8)} &
\colhead{(9)} &
\colhead{(10)} &
\colhead{(11)} &
\colhead{(12)}
}
\startdata
1 & 073106.85+392644.5 & 112.7786 & 39.44571 & 0.048306 & 29.97 & 1.032 & 14 & 17.90 & 0.55 & 5.53 & 5.36 \\
2 & 073714.28+292634.2 & 114.3095 & 29.44284 & 0.079642 & 38.90 & 1.217 & 15 & 10.85 & 0.41 & 3.43 & 2.82 \\
3 & 075525.29+391109.9 & 118.8554 & 39.18609 & 0.033165 & 51.02 & 1.175 & 11 & 26.86 & 0.51 & 7.36 & 6.27 \\
4 & 081324.01+542236.7 & 123.3501 & 54.37688 & 0.042641 & 47.85 & 1.129 & 8  & 16.28 & 0.46 & 4.57 & 4.05 \\
5 & 081345.89+381049.6 & 123.4412 & 38.18047 & 0.079993 & 23.91 & 1.144 & 12 & 10.33 & 0.39 & 3.12 & 2.73 \\
\enddata

\tablecomments{
Only the first five rows are shown here. The complete table of 101 barred NLS1s is available in a CSV form.
Column (1) gives the pair identifier, which links each barred galaxy to its matched unbarred counterpart in Table~\ref{tab:unbarred_sample}.
Column (2) gives the SDSS source name.
Columns (3) and (4) give the right ascension and declination in degrees, respectively.
Column (5) gives the spectroscopic redshift.
Column (6) gives the host-galaxy inclination angle derived from the outer-isophote axis ratio.
Column (7) gives the effective PSF FWHM of the PS1 $r$-band image.
Column (8) gives the number of PS1 $r$-band stack exposures, $N_r$.
Column (9) gives $R_{25.5}$, defined as the semi-major axis at which the PS1 $r$-band surface brightness reaches 25.5 mag arcsec$^{-2}$.
Column (10) gives the maximum bar ellipticity, $e_{\max}$.
Column (11) gives the bar length, $a_{\rm bar}$, defined as the semi-major axis at $e_{\max}$.
Column (12) gives the bar length normalized by the effective PSF FWHM, $a_{\rm bar}/\mathrm{FWHM}_{\rm eff}$.}
\end{deluxetable*}

\begin{deluxetable*}{lcccccccc}
\tablecaption{Properties of the Matched Unbarred NLS1 Sample
\label{tab:unbarred_sample}}

\tablehead{
\colhead{Pair} &
\colhead{SDSS Name} &
\colhead{R.A.} &
\colhead{Decl.} &
\colhead{$z$} &
\colhead{$i$} &
\colhead{$\mathrm{FWHM}_{\mathrm{eff}}$} &
\colhead{$N_r$} &
\colhead{$R_{25.5}$} \\
\colhead{} &
\colhead{} &
\colhead{(deg)} &
\colhead{(deg)} &
\colhead{} &
\colhead{(deg)} &
\colhead{(arcsec)} &
\colhead{} &
\colhead{(arcsec)} \\
\colhead{(1)} &
\colhead{(2)} &
\colhead{(3)} &
\colhead{(4)} &
\colhead{(5)} &
\colhead{(6)} &
\colhead{(7)} &
\colhead{(8)} &
\colhead{(9)}
}

\startdata
1 & 162409.37+260430.9 & 246.03910 &  26.07525 & 0.039237 & 27.93 & 0.994 &  8 & 16.81 \\
2 & 141755.54+431155.9 & 214.48140 &  43.19886 & 0.078909 & 38.42 & 1.036 & 10 &  7.08 \\
3 & 161809.37+361957.8 & 244.53910 &  36.33273 & 0.033763 & 48.87 & 0.976 & 13 & 12.47 \\
4 & 102531.28+514034.8 & 156.38030 &  51.67635 & 0.044586 & 50.73 & 1.038 & 15 & 15.03 \\
5 & 012755.93-103844.2 &  21.98308 & -10.64563 & 0.082405 & 24.87 & 1.087 & 15 & 11.44 \\
\enddata

\tablecomments{
Only the first five rows are shown here. The complete table of 101 unbarred NLS1s is available in a CSV form.
Column (1) gives the pair identifier, which links each unbarred galaxy to its matched barred counterpart in Table~\ref{tab:barred_sample}.
An asterisk following the pair identifier denotes an unbarred galaxy with an E/S0-like or ambiguous morphology in our conservative visual morphological check. 
These objects, together with their corresponding barred counterparts, are excluded from the clean disk-galaxy subsample discussed in Section~\ref{sec:Dis}.
Column (2) gives the SDSS source name.
Columns (3) and (4) give the right ascension and declination in degrees,
respectively.
Column (5) gives the spectroscopic redshift.
Column (6) gives the host-galaxy inclination angle derived from the
outer-isophote axis ratio.
Column (7) gives the effective PSF FWHM of the PS1 $r$-band image.
Column (8) gives the number of PS1 $r$-band stack exposures, $N_r$.
Column (9) gives $R_{25.5}$, defined as the semi-major axis at which
the PS1 $r$-band surface brightness reaches
25.5 mag arcsec$^{-2}$.
}
\end{deluxetable*}

\begin{table*}[ht]
\centering
\caption{Comprehensive Statistical Comparison between Barred and Unbarred NLS1s}
\label{tab:full_stats}
\scriptsize
\begin{tabular}{lcccc}
\hline
\hline
Property & Barred & Unbarred & Sample Size & Stat. Significance \\
& Mean $\pm$ Std & Mean $\pm$ Std & ($N_{\rm B}, N_{\rm U}$) & $p_{\mathrm{KS}}$ / $p_{\mathrm{Welch}}$ \\
(1) & (2) & (3) & (4) & (5) \\
\hline

\multicolumn{5}{c}{\textit{Basic Properties}} \\
Redshift
& $0.0644 \pm 0.0207$
& $0.0650 \pm 0.0199$
& (101, 101)
& 0.970 / 0.835 \\

Inclination (deg)
& $36.66 \pm 12.11$
& $37.13 \pm 11.50$
& (101, 101)
& 0.995 / 0.777 \\

PS1 $r$-band FWHM (arcsec)
& $1.134 \pm 0.104$
& $1.142 \pm 0.107$
& (101, 101)
& 0.912 / 0.569 \\

PS1 $N_r$
& $14.34 \pm 4.43$
& $14.66 \pm 4.28$
& (101, 101)
& 0.912 / 0.595 \\

\hline
\multicolumn{5}{c}{\textit{Continuum and Broad-Line Luminosities ($\log$ erg s$^{-1}$)}} \\

$\log[\lambda L_{\lambda}(5100~\mbox{\AA})]$
& $42.80 \pm 0.43$
& $42.91 \pm 0.52$
& (101, 101)
& 0.081 / 0.089 \\

$\log L_{\mathrm{H}\alpha,\mathrm{broad}}$
& $41.49 \pm 0.39$
& $41.63 \pm 0.51$
& (101, 101)
& 0.010 / 0.024 \\

$\log L_{\mathrm{H}\beta,\mathrm{broad}}$
& $40.94 \pm 0.42$
& $41.06 \pm 0.55$
& (101, 101)
& 0.038 / 0.078 \\

\hline
\multicolumn{5}{c}{\textit{Narrow-Line Luminosities ($\log$ erg s$^{-1}$)}} \\

$\log L_{\mathrm{H}\alpha,\mathrm{narrow}}$
& $40.43 \pm 0.56$
& $40.62 \pm 0.56$
& (95, 98)
& 0.023 / 0.015 \\

$\log L_{\mathrm{H}\beta,\mathrm{narrow}}$
& $39.63 \pm 0.59$
& $39.87 \pm 0.60$
& (94, 97)
& 0.0098 / 0.0068 \\

$\log L([\mathrm{O\,III}]~\lambda 5007)$
& $40.52 \pm 0.41$
& $40.69 \pm 0.45$
& (101, 101)
& 0.0065 / 0.0047 \\

$\log L([\mathrm{N\,II}]~\lambda 6585)$
& $40.49 \pm 0.43$
& $40.56 \pm 0.53$
& (98, 98)
& 0.105 / 0.343 \\

\hline
\multicolumn{5}{c}{\textit{Line Ratios and Gas Diagnostics}} \\

N2 $\equiv \log([\mathrm{N\,II}]~\lambda 6585/\mathrm{H}\alpha)$
& $0.077 \pm 0.318$
& $-0.058 \pm 0.288$
& (89, 92)
& 0.011 / 0.003 \\

O3Hb $\equiv \log([\mathrm{O\,III}]~\lambda 5007/\mathrm{H}\beta)$
& $0.910 \pm 0.547$
& $0.850 \pm 0.485$
& (89, 92)
& 0.900 / 0.431 \\

O3N2
& $0.833 \pm 0.520$
& $0.907 \pm 0.511$
& (89, 92)
& 0.340 / 0.336 \\

$[\mathrm{S\,II}]~\lambda 6716/\lambda 6731$
& $1.183 \pm 0.162$
& $1.164 \pm 0.162$
& (96, 91)
& 0.322 / 0.436 \\

$n_e([\mathrm{S\,II}])$ (cm$^{-3}$)
& $408 \pm 365$
& $497 \pm 714$
& (93, 87)
& 0.248 / 0.297 \\

\hline
\multicolumn{5}{c}{\textit{Black Hole and Accretion Properties}} \\

$\log M_{\rm BH}$ ($M_\odot$)
& $6.25 \pm 0.39$
& $6.34 \pm 0.40$
& (101, 101)
& 0.081 / 0.112 \\

$\log \lambda_{\rm Edd}$
& $-0.625 \pm 0.442$
& $-0.598 \pm 0.433$
& (101, 101)
& 0.970 / 0.668 \\

$\log \mathrm{FWHM}_{\mathrm{H}\alpha,\mathrm{broad}}$
& $3.101 \pm 0.138$
& $3.125 \pm 0.163$
& (101, 101)
& 0.820 / 0.252 \\

$\log \mathrm{FWHM}_{\mathrm{H}\beta,\mathrm{broad}}$
& $3.107 \pm 0.161$
& $3.110 \pm 0.139$
& (101, 101)
& 0.820 / 0.906 \\

\hline
\multicolumn{5}{c}{\textit{Extinction, Fe II Strength, and Radio Properties}} \\

$\log(\mathrm{H}\alpha/\mathrm{H}\beta)_{\rm narrow}$
& $0.734 \pm 0.336$
& $0.677 \pm 0.371$
& (83, 87)
& 0.253 / 0.292 \\

$R_{4570}$
& $0.685 \pm 0.327$
& $0.595 \pm 0.351$
& (101, 101)
& 0.159 / 0.060 \\

$\log R_{\rm L}$
& $0.637 \pm 0.367$
& $0.426 \pm 0.464$
& (16, 30)
& 0.079 / 0.099 \\

\hline
\end{tabular}
\tablecomments{Column (1) lists the galaxy or AGN property being compared. Columns (2) and (3) give the mean and standard deviation for the barred and unbarred NLS1 subsamples, respectively. Column (4) gives the numbers of barred and unbarred objects used in each comparison. Column (5) gives the $p$-values from the two-sample Kolmogorov--Smirnov test and the independent Welch $t$-test, respectively.
Logarithmic quantities are in base 10.}
\label{tab:tab1}
\end{table*}

\subsection{Nuclear Continuum and Broad-Line Emission}

In \citet{Paliya2024}, the AGN and host-galaxy starlight were decomposed using the \texttt{BADASS} package \citep{Sexton2021,Sexton2022} to obtain both emission-line and continuum properties. The spectra were shifted to the rest frame, corrected for Galactic extinction, and modeled with a combination of stellar templates, AGN continuum, and emission-line components using a Bayesian Markov Chain Monte Carlo framework \citep{MCMC2013}.

\begin{figure}[ht!]
  \centering
  \includegraphics[width=\columnwidth]{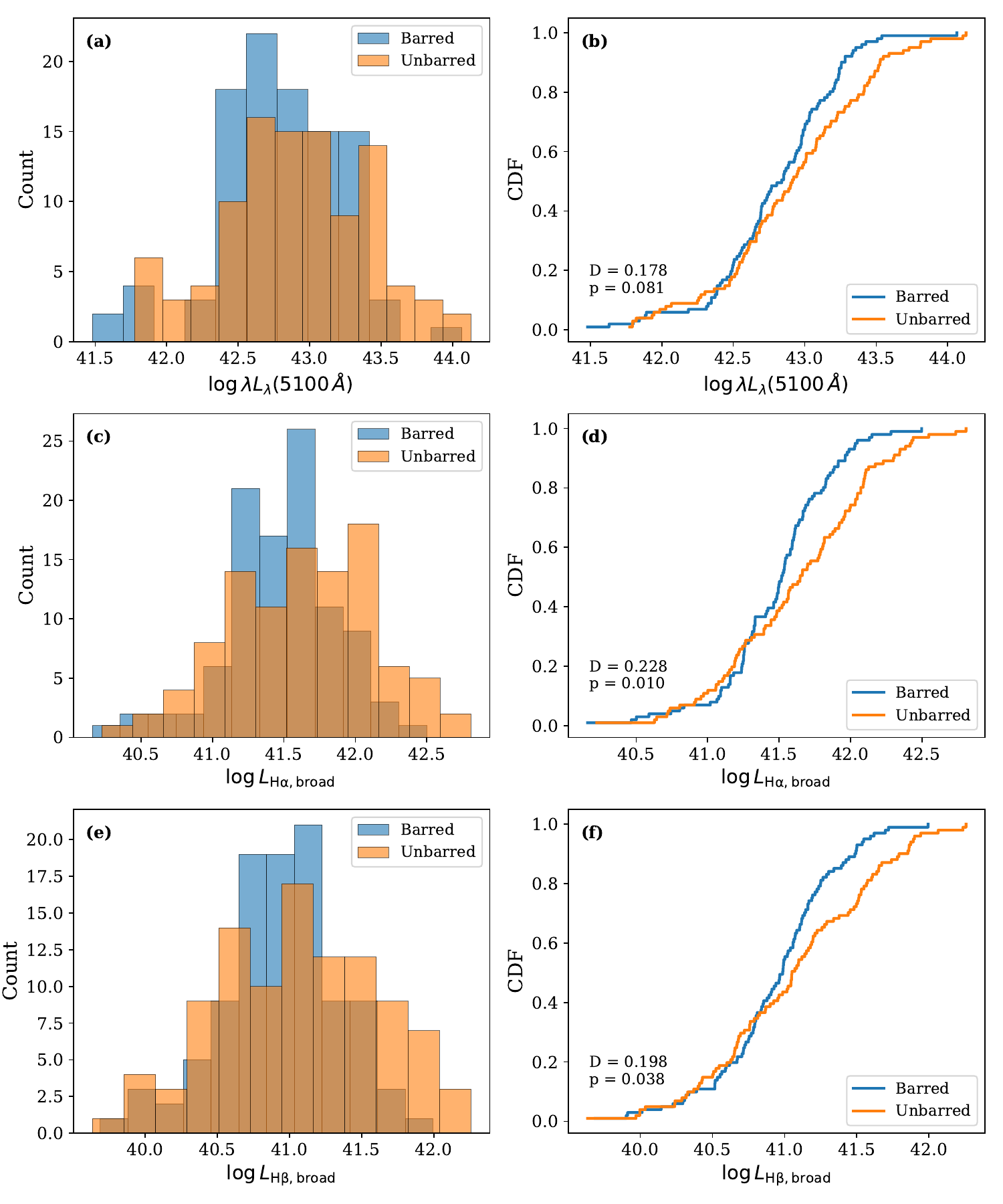}
  \caption{Distributions of the nuclear continuum and broad-line luminosities for the barred and unbarred NLS1 subsamples. The left panels show histograms, while the right panels show the corresponding cumulative distribution functions (CDFs). Panels (a) and (b) present the AGN continuum luminosity $\lambda L_{\lambda}(5100\,\text{\AA})$, panels (c) and (d) show the broad H$\alpha$ luminosity, and panels (e) and (f) show the broad H$\beta$ luminosity. The Kolmogorov–Smirnov test statistics are indicated in the CDF panels.}
\label{fig:fig5}
\end{figure}

The monochromatic luminosity at 5100~\mbox{\AA}, denoted as $\lambda L_{\lambda}(5100~\mbox{\AA})$, is commonly used as a tracer of the
optical AGN continuum \citep{Kaspi2000,Vestergaard2006}, as this spectral region is relatively free from strong emission-line contamination. The unbarred NLS1s show a modestly higher mean continuum luminosity than the barred NLS1s, with $\langle \log[\lambda L_{\lambda}(5100~\mbox{\AA})]\rangle = 42.91 \pm 0.52$ for the unbarred subsample and $42.80 \pm 0.43$ for the barred subsample. However, the difference is only marginal in the two-sample tests, with $p_{\mathrm{KS}}=0.081$ and $p_{\mathrm{Welch}}=0.089$ (Figure~\ref{fig:fig5}). 

We also assessed the robustness of the 5100~\mbox{\AA} continuum offset using Monte Carlo resampling with measurement uncertainties and the
Bayesian model. The Monte Carlo analysis gives a positive mean offset of $\Delta = 0.115 \pm 0.011$ dex, where $\Delta=\langle x\rangle_{\mathrm{unbarred}}-\langle x\rangle_{\mathrm{barred}}$, with all realizations yielding $\Delta>0$. The Bayesian model gives $\Delta = 0.117 \pm 0.067$ dex and $P(\Delta>0)=0.961$. These results indicate that the direction of the continuum offset is robust, although its statistical significance remains modest.

We further compared the luminosities of the broad H$\alpha$ and H$\beta$ components between the barred and unbarred NLS1s (Figure~\ref{fig:fig5}). The broad H$\alpha$ luminosity shows a clearer offset: the barred NLS1s have $\langle \log L_{\mathrm{H}\alpha,\mathrm{broad}}\rangle = 41.49 \pm 0.39$, while the unbarred NLS1s have a higher mean value of $41.63 \pm 0.51$. This difference is significant in both the K--S and Welch tests, with $p_{\mathrm{KS}}=0.010$ and
$p_{\mathrm{Welch}}=0.024$. The broad H$\beta$ luminosity shows the same direction of offset, with mean values of $\langle \log L_{\mathrm{H}\beta,\mathrm{broad}}\rangle = 40.94 \pm 0.42$ for barred NLS1s and $41.06 \pm 0.55$ for unbarred NLS1s. The K--S test indicates a difference between the two distributions ($p_{\mathrm{KS}}=0.038$), while the Welch test gives a marginal result ($p_{\mathrm{Welch}}=0.078$).

The Monte Carlo and Bayesian analyses support the same overall trend. For broad H$\alpha$, the Monte Carlo resampling gives $\Delta = 0.145 \pm 0.001$ dex, while the Bayesian model gives $\Delta = 0.145 \pm 0.064$ dex with $P(\Delta>0)=0.988$. For broad H$\beta$, the corresponding offsets are $\Delta = 0.122 \pm 0.003$ dex from the Monte Carlo analysis and $\Delta = 0.122 \pm 0.069$ dex from the Bayesian model, with $P(\Delta>0)=0.963$. Thus, the broad-line results are consistent with the continuum comparison: unbarred NLS1s are modestly but systematically brighter than barred NLS1s in optical AGN luminosity indicators.

\subsection{Narrow-Line Luminosities}

We also compared the narrow-line luminosities between the barred and
unbarred subsamples. The unbarred NLS1s are systematically brighter in the
main narrow recombination lines. For narrow H$\alpha$, the barred subsample
has $\langle \log L_{\mathrm{H}\alpha,\mathrm{narrow}}\rangle = 40.43 \pm 0.56$, while the unbarred subsample has
$40.62 \pm 0.56$. This difference is significant in both the K--S and Welch tests, with $p_{\mathrm{KS}}=0.023$ and
$p_{\mathrm{Welch}}=0.015$. For narrow H$\beta$, the barred and unbarred mean luminosities are
$\langle \log L_{\mathrm{H}\beta,\mathrm{narrow}}\rangle = 39.63 \pm 0.59$ and $39.87 \pm 0.60$, respectively, with
$p_{\mathrm{KS}}=0.0098$ and $p_{\mathrm{Welch}}=0.0068$.

The Monte Carlo resampling and the Bayesian model support the same direction of offset. For narrow H$\alpha$,
the Monte Carlo analysis gives $\Delta = 0.197 \pm 0.013$ dex, while the Bayesian model gives $\Delta = 0.183 \pm 0.078$ dex with $P(\Delta>0)=0.991$. 
For narrow H$\beta$, the corresponding offsets are $\Delta = 0.236 \pm 0.055$ dex from the Monte Carlo analysis and $\Delta = 0.221 \pm 0.086$ dex from the Bayesian model, with $P(\Delta>0)=0.995$. Here $\Delta=\langle x\rangle_{\mathrm{unbarred}}-\langle x\rangle_{\mathrm{barred}}$.

The [\ion{O}{3}]~$\lambda 5007$ emission line is commonly used as a tracer of NLR emission associated with AGN photoionization \citep{{Kauffmann2003,Heckman2005}}, although it can also be affected by NLR gas content, covering factor, density, and extinction. In the PS1-resolved matched sample, the unbarred NLS1s show higher [\ion{O}{3}] luminosities than the barred NLS1s. The barred subsample has a mean $\log L([\mathrm{O\,III}]~\lambda 5007) = 40.52 \pm 0.41$, while the unbarred subsample has a higher mean value of $40.69 \pm 0.45$. The two-sample tests indicate that this difference is significant, with $p_{\mathrm{KS}}=0.0065$ and $p_{\mathrm{Welch}}=0.0047$. The Monte Carlo analysis gives a positive offset of $\Delta = 0.172 \pm 0.005$ dex, and the Bayesian model gives $\Delta = 0.171 \pm 0.061$ dex with $P(\Delta>0)=0.998$. This result is consistent with the broader trend that barred NLS1s are not enhanced in optical AGN luminosity.

The [\ion{N}{2}]~$\lambda 6585$ line is a low-ionization forbidden transition that can arise from both the NLR and other ionized gas components.
Unlike [\ion{O}{3}], its strength depends not only on the AGN ionizing radiation, but also on gas-phase metallicity and local ionization conditions.
In contrast to the [\ion{O}{3}] result, the [\ion{N}{2}] luminosity does not show a significant barred--unbarred difference. The barred and unbarred mean
values are $\log L([\mathrm{N\,II}]~\lambda 6585)=40.49 \pm 0.43$ and $40.56 \pm 0.53$, respectively, with $p_{\mathrm{KS}}=0.105$ and $p_{\mathrm{Welch}}=0.343$.

Overall, the narrow-line results show that unbarred NLS1s are brighter in narrow H$\alpha$, narrow H$\beta$, and [\ion{O}{3}] emission, whereas
[\ion{N}{2}] luminosity remains statistically comparable between the two subsamples.

\subsection{Ionization and Physical Conditions of the Narrow-Line Regions}
\label{subsec:nlr_conditions}

\begin{figure}[ht!]
  \centering
  \includegraphics[width=\columnwidth]{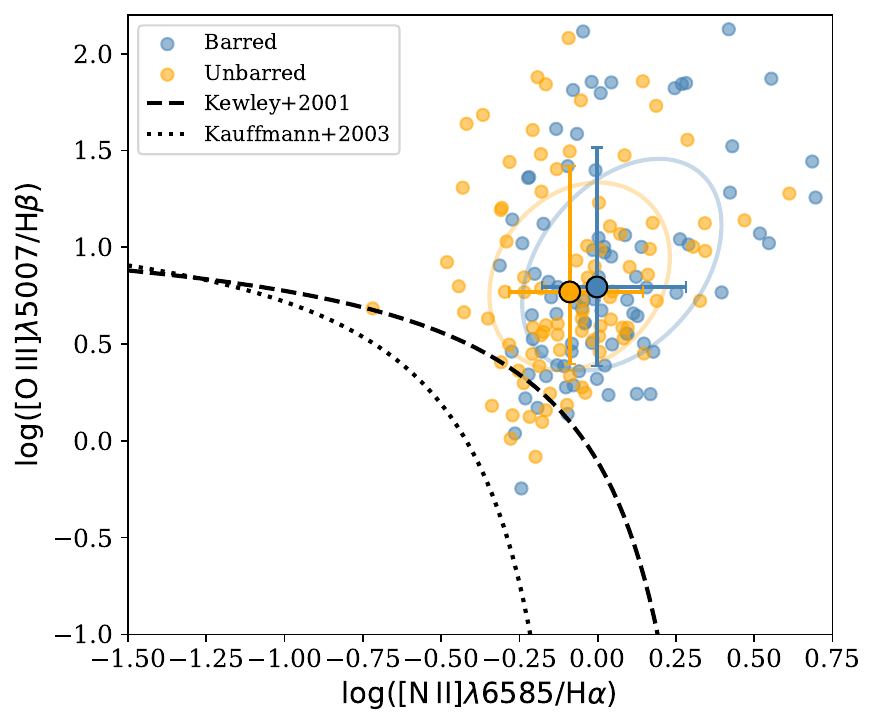}
  \caption{The BPT diagram for barred (blue) and unbarred (orange) NLS1 galaxies. Both populations occupy the AGN region with no obvious separation in $\log([\mathrm{O\,III}]\,\lambda 5007/\mathrm{H}\beta)$, although barred systems show a mild shift toward higher $\log([\mathrm{N\,II}]\,\lambda 6585/\mathrm{H}\alpha)$. The dashed and dotted curves show the demarcation lines from \citet{Kewley2001} and \citet{Kauffmann2003}, respectively.}
 \label{fig:fig6}
\end{figure}

We examined several narrow-line diagnostic ratios to assess whether barred and unbarred NLS1s differ in the physical conditions of their NLRs.
Figure~\ref{fig:fig6} presents the Baldwin--Phillips--Terlevich (BPT) diagram \citep{Baldwin1981} for the barred and unbarred NLS1 subsamples. Requiring valid measurements for [\ion{O}{3}], H$\beta$, [\ion{N}{2}], and H$\alpha$ leaves 89 barred and 92 unbarred NLS1s. Both populations lie predominantly above the \citet{Kewley2001} demarcation line, indicating that the narrow-line emission is dominated by AGN photoionization rather than star formation.

The ionization-sensitive ratio $\log([\mathrm{O\,III}]~\lambda 5007/\mathrm{H}\beta)$ shows no significant difference between the two subsamples. The barred and unbarred mean values are $0.910 \pm 0.547$ and $0.850 \pm 0.485$, respectively, with $p_{\mathrm{KS}}=0.900$ and $p_{\mathrm{Welch}}=0.431$. This indicates that the characteristic NLR ionization state, as traced by [\ion{O}{3}]/H$\beta$, is broadly similar in barred and unbarred systems.

A clearer offset is seen in the N2 index, defined as $\log([\mathrm{N\,II}]~\lambda 6585/\mathrm{H}\alpha)$. The barred subsample has a higher mean N2 value of $0.077 \pm 0.318$, compared with $-0.058 \pm 0.288$ for the unbarred subsample. This difference is significant in both the K--S and Welch tests, with $p_{\mathrm{KS}}=0.011$ and $p_{\mathrm{Welch}}=0.003$. Thus, barred NLS1s show relatively stronger low-ionization nitrogen emission with respect to narrow H$\alpha$.

The O3N2 index, defined here as $\log([\mathrm{O\,III}]~\lambda 5007/\mathrm{H}\beta)-\log([\mathrm{N\,II}]~\lambda 6585/\mathrm{H}\alpha)$, compares the relative strength of high-ionization and low-ionization narrow-line emission. The mean O3N2 values are $0.833 \pm 0.520$ for barred NLS1s and $0.907 \pm 0.511$ for unbarred NLS1s. Although the unbarred subsample has a slightly higher mean O3N2 value, the difference is not statistically significant, with $p_{\mathrm{KS}}=0.340$ and $p_{\mathrm{Welch}}=0.336$. Therefore, the N2 offset is not accompanied by a significant separation in O3N2.

We also used the [\ion{S}{2}] doublet ratio, $[\mathrm{S\,II}]~\lambda 6716/\lambda 6731$, as a density diagnostic. The barred and unbarred subsamples show statistically consistent [\ion{S}{2}] ratios, with mean values of $1.183 \pm 0.162$ and $1.164 \pm 0.162$, respectively, where the quoted uncertainties represent the standard deviations of the distributions ($p_{\mathrm{KS}}=0.322$, $p_{\mathrm{Welch}}=0.436$). Using PyNeb and assuming $T_e=10^4~\mathrm{K}$, the corresponding electron densities are $n_e=408 \pm 365~\mathrm{cm}^{-3}$ for barred NLS1s and $497 \pm 714~\mathrm{cm}^{-3}$ for unbarred NLS1s. These densities are not significantly different ($p_{\mathrm{KS}}=0.248$, $p_{\mathrm{Welch}}=0.297$). The inferred electron density depends only weakly on $T_e$ over the temperature range typical of ionized nebulae \citep{Copetti2000,Li2025}. In particular, variations in $T_e$ from approximately 5000 to 20,000 K change the inferred $n_e$ by $\lesssim0.2$ dex \citep{Li2025}. Because the same temperature assumption is applied consistently to both subsamples, this choice is unlikely to significantly affect the relative electron-density comparison.

Taken together, these diagnostic ratios indicate that barred and unbarred NLS1s have broadly similar NLR ionization states and electron densities. The main exception is the N2 ratio, which is higher in barred systems. However, this difference is not accompanied by a significant enhancement of the [\ion{N}{2}] luminosity itself. As shown in the narrow-line luminosity comparison, [\ion{N}{2}] luminosity remains statistically comparable between
the two subsamples, whereas narrow H$\alpha$ is lower in barred NLS1s. The higher N2 ratio in barred systems therefore reflects the different relative behavior of [\ion{N}{2}] and lower H$\alpha$, rather than a simple enhancement of [\ion{N}{2}] emission.

\subsection{Dependence on Bar Strength}
\label{subsec:bar_strength}

\begin{figure*}[ht!]
  \centering
  \includegraphics[width=\textwidth]{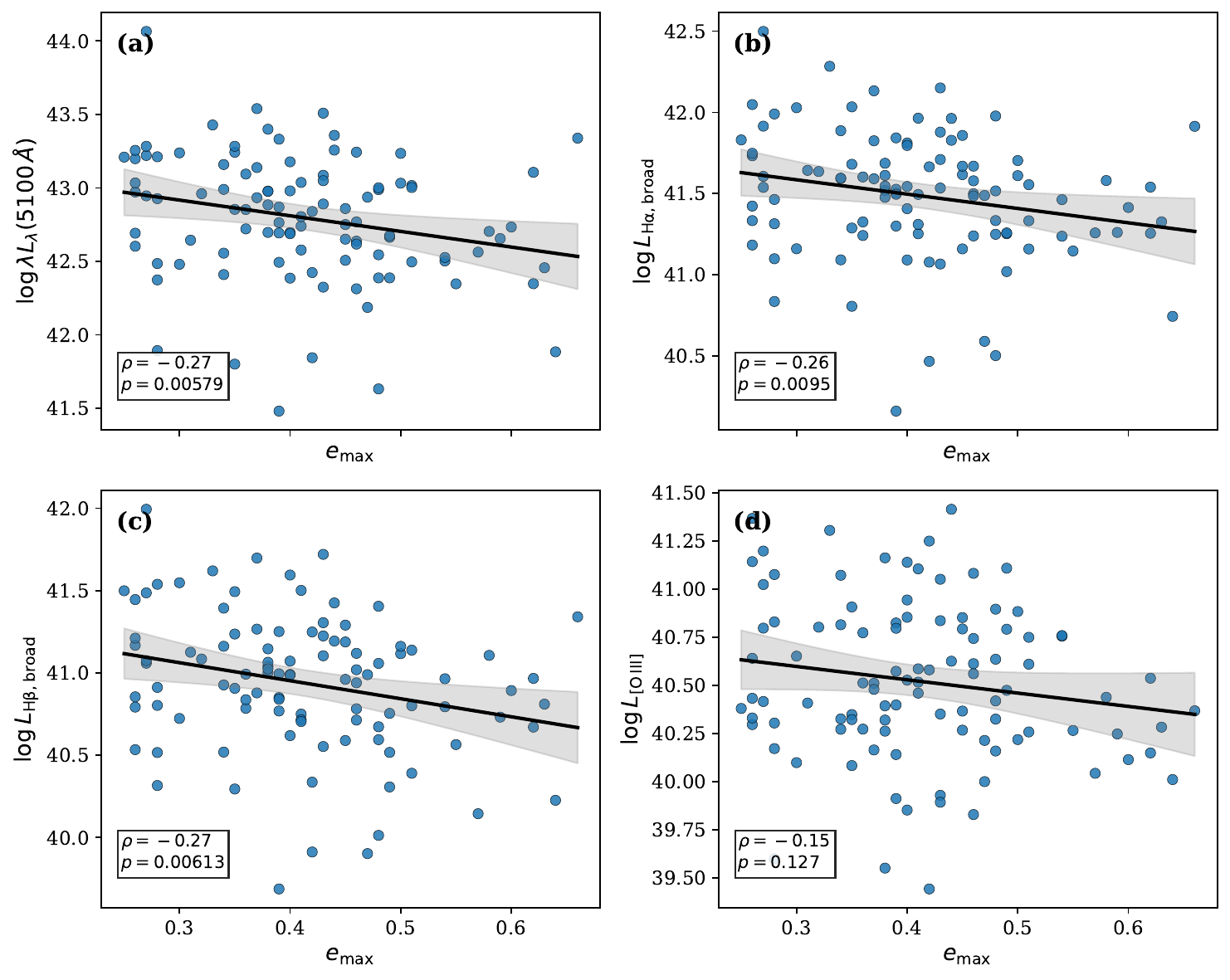}
  \caption{Relation between bar strength ($e_{\rm max}$) and AGN luminosity indicators for the barred NLS1 subsample: 
(a) $\lambda L_{\lambda}(5100\,\AA)$, (b) broad H$\alpha$ luminosity, (c) broad H$\beta$ luminosity, and (d) [O\,{\sc iii}] luminosity.
The black lines show the best-fit linear relations and the shaded regions indicate the 95\% confidence intervals.}
\label{fig:fig7}
\end{figure*}

We investigated whether the apparent stellar-bar strength, quantified by $e_{\max}$, correlates with AGN luminosity indicators within the barred
subsample. Bar ellipticity is commonly used as a photometric proxy for bar strength \citep{Marinova2007,Aguerri2009}, although it does not directly measure the gravitational bar torque. For the full barred sample ($N=101$), $e_{\max}$ shows a mild but statistically significant anti-correlation with the optical continuum
luminosity $\lambda L_{\lambda}(5100~\mbox{\AA})$ (Figure~\ref{fig:fig7}). The Spearman rank coefficient is $\rho=-0.273$ with $p=0.0058$, while the Pearson test gives $r=-0.247$ with $p=0.0128$.

Similar anti-correlations are found for the broad Balmer-line luminosities. For broad H$\alpha$, the Spearman coefficient is
$\rho=-0.257$ with $p=0.0095$, and the Pearson coefficient is $r=-0.226$ with $p=0.023$. For broad H$\beta$, the corresponding values are
$\rho=-0.271$ with $p=0.0061$ and $r=-0.260$ with $p=0.0086$. In contrast, no statistically significant correlation is found between
$e_{\max}$ and the narrow-line [\ion{O}{3}]~$\lambda 5007$ luminosity ($\rho=-0.153$, $p=0.127$).

To examine possible projection effects, we divided the barred sample into inclination bins of $0^\circ$--$20^\circ$, $20^\circ$--$40^\circ$, and
$40^\circ$--$60^\circ$. The lowest-inclination bin contains only eight objects and is therefore too small for a meaningful statistical
interpretation. In the intermediate inclination range ($20^\circ$--$40^\circ$, $N=52$), the anti-correlation between $e_{\max}$
and $\lambda L_{\lambda}(5100~\mbox{\AA})$ remains significant, with $\rho=-0.313$ and $p=0.0237$; the Pearson test gives
$r=-0.320$ and $p=0.0209$. The broad-line luminosities show similar negative trends in this bin, although they do not reach formal statistical
significance. In the higher-inclination bin ($40^\circ$--$60^\circ$, $N=41$), the correlations remain negative but are statistically marginal,
with $\rho=-0.276$ and $p=0.080$ for the continuum luminosity.

These results indicate that the apparent bar-strength--continuum anti-correlation is present in the full barred sample and is most clearly
detected in systems with moderate inclination. The trend weakens in smaller inclination subsamples and at higher inclination, where projection effects
and deprojection uncertainties can increase the scatter in the measured bar ellipticity. Thus, stronger observed bar signatures are not associated with enhanced AGN luminosity. The result is consistent with \citet{SilvaLima2022}.

As an alternative indicator of bar prominence, we defined $a_{\rm bar}$ as the SMA at $e_{\max}$ and examined $a_{\rm bar}/{\rm FWHM}_{\rm eff}$ within the barred subsample. No statistically significant correlations are found between $a_{\rm bar}/{\rm FWHM}_{\rm eff}$ and the AGN luminosity indicators,
including $\lambda L_\lambda(5100\,\text{\AA})$, broad H$\alpha$, broad H$\beta$, and [O~III] luminosities; all Spearman correlation $p$-values are greater than 0.3. Although the weak negative correlations obtained with $e_{\max}$ are not reproduced, the normalized-bar-length analysis likewise provides no evidence that more prominent bars are associated with enhanced AGN activity.

\subsection{Balmer Decrement and Obscuration}
\label{subsec:balmer}

\begin{figure}[ht!]
  \centering
  \includegraphics[width=\columnwidth]{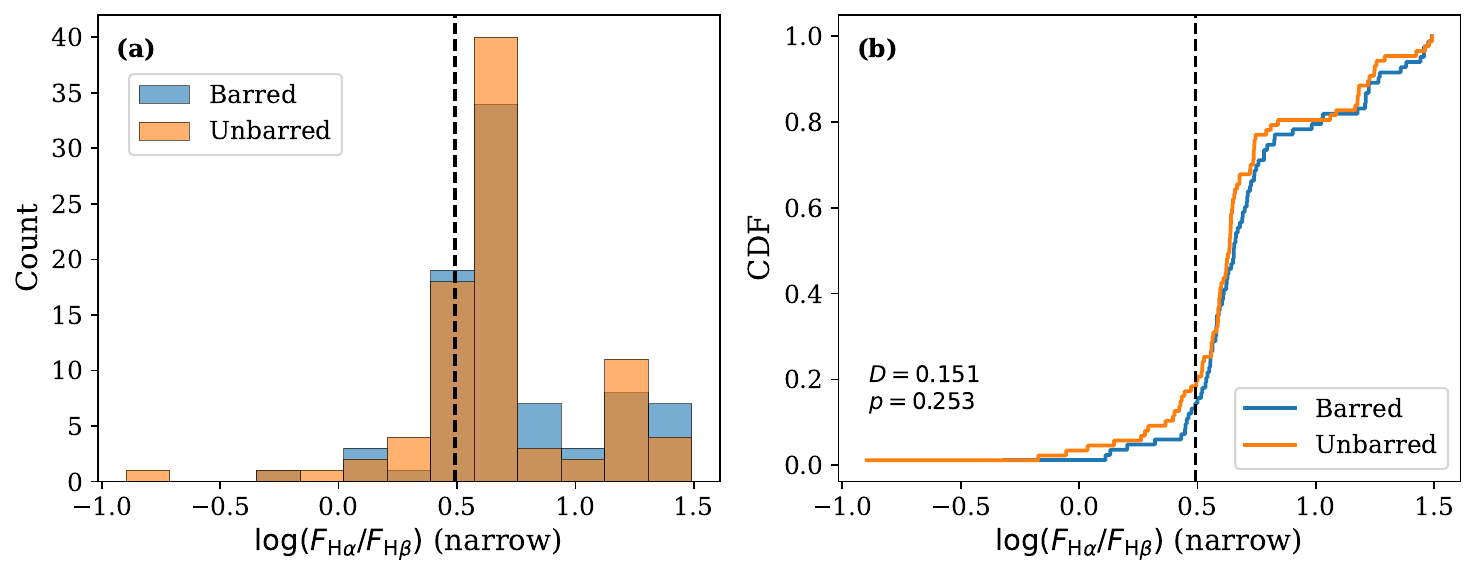}
  \caption{Distribution of the narrow-line Balmer decrement, $\log(F_{\mathrm{H}\alpha}/F_{\mathrm{H}\beta})$, for the barred and unbarred NLS1 subsamples. Panel (a) shows the histogram, and panel (b) shows the cumulative distribution function. The two subsamples are statistically indistinguishable.}
 \label{fig:fig8}
\end{figure}

To explore whether differential dust attenuation could account for the observed luminosity offsets, we compared both the narrow-line and broad-line
Balmer decrements, $\log(F_{\mathrm{H}\alpha}/F_{\mathrm{H}\beta})$, between the barred and unbarred subsamples (Figure~\ref{fig:fig8}). 

For the broad-line ratio, we treat $\log(F_{\mathrm{H}\alpha}/F_{\mathrm{H}\beta})_{\rm broad}$ as a broad-line Balmer-decrement or BLR color proxy. Although
\citet{Dong2008} showed that the intrinsic broad-line H$\alpha$/H$\beta$ ratio is tightly distributed around 3.06 (with a scatter of $\sim$0.03 dex) and largely independent of AGN properties, this intrinsic value differs from, and shows nonzero scatter relative to, the essentially fixed narrow-line Case-B ratio of 2.86 \citep{Osterbrock2006}.

For the narrow-line Balmer decrement, the barred subsample has $\langle \log(F_{\mathrm{H}\alpha}/F_{\mathrm{H}\beta})_{\rm narrow}
\rangle = 0.734 \pm 0.336$, while the unbarred subsample has a comparable mean value of $0.677 \pm 0.371$. The difference is not statistically
significant, with $p_{\mathrm{KS}}=0.253$ and $p_{\mathrm{Welch}}=0.292$.

The broad-line Balmer decrement also shows no significant barred--unbarred difference. The barred and unbarred subsamples have mean values of
$\langle \log(F_{\mathrm{H}\alpha}/F_{\mathrm{H}\beta})_{\rm broad} \rangle = 0.546 \pm 0.123$ and $0.569 \pm 0.131$, respectively. The
two-sample tests give $p_{\mathrm{KS}}=0.159$ and $p_{\mathrm{Welch}}=0.194$.

The similarity of both the narrow-line and broad-line Balmer decrements indicates no evidence for systematically stronger dust attenuation in the
barred subsample. Therefore, the systematically lower optical continuum and emission-line luminosities observed in barred NLS1s are unlikely to be
primarily caused by enhanced dust extinction in barred systems.

We also examined possible X-ray obscuration using archival X-ray source catalogs. The final optical sample was cross-matched with the Chandra Source
Catalog 2.1 \citep[CSC~2.1;][]{Evans2024} using a search radius of $3''$, and with the HEASARC XMM-Newton Serendipitous Source Catalog DR 15 \citep[5XMM-DR15;][]{Watson2009,Rosen2016,Webb2020} using a search radius of $5''$. The XMMSSC cross-match yields 12 barred and 15 unbarred detections out of 101 objects in each subsample, corresponding to detection fractions of 11.9\% and 14.9\%, respectively. The CSC~2.1 cross-match yields 18 barred and 13 unbarred detections, corresponding to detection fractions of 17.8\% and 12.9\%. These detection fractions are not significantly different between the barred and unbarred subsamples with Fisher's exact test ($p_{\rm Fisher}=0.680$ for XMMSSC and $p_{\rm Fisher}=0.435$ for CSC~2.1).

Among the detected sources, we use the catalog hardness ratios as auxiliary diagnostics of possible X-ray absorption. The XMMSSC hardness ratios show no
significant barred--unbarred differences (HR2: $p_{\rm KS}=0.990$, $p_{\rm Welch}=0.940$; HR3: $p_{\rm KS}=0.151$, $p_{\rm Welch}=0.285$). Similarly, the CSC~2.1 hardness ratios show no significant differences (HR$_{\rm HM}$: $p_{\rm KS}=0.979$, $p_{\rm Welch}=0.856$; HR$_{\rm MS}$: $p_{\rm KS}=0.355$, $p_{\rm Welch}=0.200$). Because X-ray absorption tends to harden the observed spectrum, the lack of a significant hardness-ratio difference suggests that the optical luminosity offset is unlikely to be driven primarily by systematically stronger X-ray obscuration in barred systems. However, the X-ray detected subsamples are small, so this comparison is used only as a supplementary check rather than a primary diagnostic.

Taken together, these results indicate that neither optical reddening nor X-ray obscuration differs significantly between barred and unbarred systems. Therefore, attenuation effects are unlikely to drive the observed luminosity offset. However, because the X-ray detected subsamples are small, differences in nuclear obscuration between the two populations cannot be entirely excluded.

\subsection{Black Hole Mass and Eddington Ratio}

We compared the catalog-based black hole mass ($M_{\rm BH}$) and Eddington ratio ($\lambda_{\rm Edd}$) distributions between the barred and unbarred NLS1
subsamples. The barred galaxies have a mean $\log(M_{\rm BH}/M_\odot)=6.25 \pm 0.39$, while the unbarred systems have a slightly larger mean value of $6.34 \pm 0.40$. However, this difference is not statistically significant, with $p_{\mathrm{KS}}=0.081$ and $p_{\mathrm{Welch}}=0.112$.

Similarly, the Eddington ratio distributions are nearly identical for the two populations. The barred subsample has a mean
$\log \lambda_{\rm Edd}=-0.625 \pm 0.442$, compared with $-0.598 \pm 0.433$ for the unbarred subsample. Both the K--S test
($p_{\mathrm{KS}}=0.970$) and the Welch test ($p_{\mathrm{Welch}}=0.668$) indicate no statistically significant
difference. These results show that the barred and unbarred subsamples have similar black hole masses and Eddington ratios, despite the modest luminosity
offsets seen in the optical continuum and emission-line luminosities.

Because the black hole masses are single-epoch virial estimates that depend on broad-line widths and empirical radius--luminosity relations, we also
compare the broad H$\alpha$ and broad H$\beta$ FWHM distributions (see Table~\ref{tab:tab1}). Neither line width shows a significant barred--unbarred difference. Thus, despite the modest luminosity offsets seen in the optical continuum and emission-line luminosities, the two subsamples show no evidence for significantly different virial black hole masses, Eddington ratios, or
broad-line kinematics.

\subsection{Fe II Strength}

Strong \ion{Fe}{2} emission, quantified by $R_{4570}$, is a defining property of NLS1 galaxies \citep{Osterbrock1985}. 
Here $R_{4570}$ is defined as the flux ratio of the \ion{Fe}{2}~$\lambda 4570$ blend to the broad component of H$\beta$. 
The barred subsample has a mean $R_{4570}=0.685 \pm 0.327$, while the unbarred subsample has $R_{4570}=0.595 \pm 0.351$. 
The difference is not significant in the K--S test ($p_{\mathrm{KS}}=0.159$), although the Welch test gives a marginal result ($p_{\mathrm{Welch}}=0.060$).
It should be noted that $R_{4570}$ is normalized by the broad H$\beta$ flux. Therefore, the slightly higher mean $R_{4570}$ in barred NLS1s should not be interpreted directly as evidence for intrinsically stronger Fe~\textsc{ii} emission; it may partly reflect the lower broad H$\beta$ luminosity in the barred subsample. Overall, the
$R_{4570}$ comparison does not suggest that the main luminosity offsets are driven by a strong systematic difference in BLR emission-line physics.

\subsection{Radio Loudness}
\label{subsec:RL}

We examined the radio-loudness parameter, $\log R_{\rm L}$, for the subsample with available radio information in \citet{Paliya2024}.
$R_{\rm L}$ is based on the ratio of the rest-frame flux densities at 5~GHz and 4400~\mbox{\AA} \citep{Kellermann1989}. In \citet{Paliya2024}, the 5~GHz
flux density was extrapolated from the 1.4~GHz integrated flux density by assuming a spectral index of $\alpha=0.5$, where $F_{\nu}\propto \nu^{-\alpha}$. 

Radio-loudness measurements are available for 16 barred and 30 unbarred NLS1s. The barred subsample has a slightly
higher mean value, $\langle \log R_{\rm L}\rangle = 0.637 \pm 0.367$, compared with $0.426 \pm 0.464$ for the unbarred subsample. However, this
difference is not statistically significant, with $p_{\mathrm{KS}}=0.079$ and $p_{\mathrm{Welch}}=0.099$. Using the conventional radio-loud threshold $\log R_{\rm L}>1$, the radio-loud fractions are nearly identical: $2/16$ for barred NLS1s and $4/30$ for unbarred NLS1s, with Fisher's exact test giving $p_{\rm Fisher}=1.0$.
Therefore, radio-loud objects are unlikely to drive the optical luminosity offsets discussed above, although the radio comparison is limited by the
small number of sources with available radio measurements.

\section{Discussion}
\label{sec:Dis}

\subsection{Robustness to Disk-Galaxy Selection}
To further examine possible contamination by non-disk galaxies in the unbarred sample, we performed an additional morphological check. We
fitted the surface-brightness profiles with both a single S\'ersic model and a two-component S\'ersic plus exponential-disk model. The profile fitting provided additional information on the presence of disk components, but was not used alone to determine the host morphology. We further visually inspected deep PS1 multi-band coadded
images constructed by combining the $g$, $r$, $i$, and $z$ stack images. These multi-band coadds were used only for morphological inspection and not for photometric measurements. After conservatively excluding galaxies with E/S0-like or ambiguous morphologies, together with their corresponding barred counterparts, the clean disk-galaxy sample contains 87 matched pairs.

We repeated the main AGN-activity comparisons using this clean sample. The unbarred NLS1s continue to show systematically higher luminosities in
$\lambda L_{\lambda}(5100~\mbox{\AA})$, broad H$\alpha$, broad H$\beta$, narrow H$\alpha$, narrow H$\beta$, and [\ion{O}{3}]~$\lambda 5007$, with mean offsets of approximately $0.14$--$0.26$ dex. The Welch's $t$-test gives $p=0.056$, $0.014$, $0.056$, $0.013$, $0.006$, and $0.007$, respectively. Thus, although the statistical significance of the individual indicators varies, the direction and magnitude of the luminosity differences remain consistent with those obtained from the full matched sample. The main conclusion is therefore unchanged after restricting the comparison to a conservatively selected clean disk-galaxy sample.

We also examined the dependence of AGN indicators, $\lambda L_{\lambda}(5100\,\AA)$, broad H$\alpha$ luminosity, broad H$\beta$ luminosity, and [\ion{O}{3}]~$\lambda 5007$ luminosity, on bar ellipticity using the clean sample. Among the 87 barred galaxies, $e_{\max}$ remains negatively correlated with $\log \lambda L_{\lambda}(5100\,\text{\AA})$, $\log L_{\mathrm{H}\alpha,\mathrm{broad}}$, and $\log L_{\mathrm{H}\beta,\mathrm{broad}}$, with Spearman correlation coefficients of $\rho=-0.383$, $-0.365$, and $-0.376$ and corresponding $p$-values of $2.5\times10^{-4}$, $5.0\times10^{-4}$, and $3.3\times10^{-4}$, respectively. No statistically significant correlation is found between $e_{\max}$ and $\log L_{[\mathrm{O\,III}]~\lambda 5007}$ ($\rho=-0.197$, $p=0.067$). These results are consistent with the conclusion from the full sample that larger bar ellipticity is not associated with enhanced AGN activity.

Moreover, we also used the normalized bar length $a_{\mathrm{bar}}/\mathrm{FWHM}_{\mathrm{eff}}$ to test its dependence on the AGN indicators using the clean sample. No statistically significant correlations are found between $a_{\mathrm{bar}}/\mathrm{FWHM}_{\mathrm{eff}}$ and $\log \lambda L_{\lambda}(5100\,\AA)$, $\log L_{\mathrm{H}\alpha,\mathrm{broad}}$, $\log L_{\mathrm{H}\beta,\mathrm{broad}}$, or $\log L_{[\mathrm{O\,III}]~\lambda 5007}$, with Spearman correlation coefficients of $\rho=-0.019$, $0.021$, $0.033$, and $-0.097$, and corresponding $p$-values of $0.863$, $0.846$, $0.762$, and $0.372$, respectively. Thus, while bar ellipticity and normalized bar length trace different structural properties of the bar, neither analysis provides evidence that more prominent bars are associated with enhanced AGN activity. These results are consistent with the conclusion obtained from the full sample.

\subsection{SDSS Fiber-Aperture Effects on the Narrow-line Comparison}

We define the normalized fiber radius as \(1\farcs5/R_{25.5}\) to examine the difference of four narrow lines, including narrow H$\alpha$, narrow H$\beta$, and [\ion{O}{3}]~$\lambda 5007$, and [\ion{N}{2}]~$\lambda 6585$ lines. Although the barred galaxies have systematically smaller normalized fiber radii than the unbarred galaxies, we find no significant correlation between this quantity and the luminosity of the narrow lines within either subsample (all Spearman $p$$>$0.1). We further adopted 
multiple linear regression analyses using \texttt{statsmodels} package \citep{Seabold2010}, with line luminosity as the dependent variable and bar classification, $\log_{10}(1\farcs5/R_{25.5})$, redshift, and inclination as predictors. The coefficient associated with bar classification remains negative and statistically significant for all four narrow lines after accounting for these quantities. Thus, the observed narrow-line differences are unlikely to be primarily driven by the different relative fiber coverage of the two subsamples.

\subsection{A Systematic Optical Luminosity Offset}

The primary result of this work is that barred NLS1s do not show enhanced optical AGN luminosity relative to their unbarred counterparts. Instead, the unbarred subsample is modestly but consistently brighter in several optical AGN luminosity indicators, including $\lambda L_{\lambda}(5100~\mbox{\AA})$, broad H$\alpha$, broad H$\beta$, narrow H$\alpha$, narrow H$\beta$, and [\ion{O}{3}]~$\lambda 5007$, with typical offsets of $\sim 0.1$--$0.25$ dex. If the offset were seen only in narrow lines, it could be affected by NLR covering factor, density structure, gas geometry, or the ionizing photon flux intercepted by the NLR gas.
However, the same sense of offset is also present in the optical continuum and broad Balmer lines. Since the BLR is much closer to the central engine \citep[e.g.,][]{Peterson1993,Osterbrock2006}, and is not governed by large-scale NLR gas geometry, the combined continuum, BLR, and NLR behavior suggests a coherent offset in the overall optical AGN
luminosity scale, rather than an effect confined to the NLR.

This luminosity offset is not accompanied by a comparable change in the main narrow-line diagnostic ratios. In particular, the
$\log([\mathrm{O\,III}]~\lambda 5007/\mathrm{H}\beta)$ ratio is statistically consistent between the barred and unbarred subsamples, with mean values of
$0.910 \pm 0.547$ and $0.850 \pm 0.485$, respectively ($p_{\mathrm{KS}}=0.900$ and $p_{\mathrm{Welch}}=0.431$). Thus, although both
narrow H$\beta$ and [\ion{O}{3}] luminosities are lower in barred NLS1s, their ratio does not show evidence for a significant barred--unbarred
difference. The barred and unbarred systems also occupy broadly similar positions in the BPT diagram, and the [\ion{S}{2}] density diagnostic shows
no significant difference between the two subsamples. These results suggest that the narrow-line luminosity offset is mainly a normalization difference,
rather than a clear change in the characteristic NLR ionization state or electron density.

The luminosity offset is also unlikely to be caused primarily by differential dust attenuation or by other obvious secondary effects. Both the narrow-line
and broad-line Balmer decrements show no significant barred--unbarred differences. Similarly, the black hole mass, Eddington ratio, radio-loud
fraction, and archival X-ray hardness ratios do not show evidence for a systematic difference capable of explaining the optical luminosity offset.
Taken together, these results indicate that large-scale stellar bars are not associated with enhanced instantaneous optical AGN luminosity in our matched NLS1 sample.

\subsection{Possible Inner Transport Bottlenecks in Barred Systems}

Large-scale stellar bars can efficiently redistribute angular momentum and drive gas toward the central kiloparsec.
Observationally, barred galaxies show enhanced central molecular gas concentrations, with stronger bars generally associated with higher central gas concentrations \citep{Kuno2007,Yu2022}. However, this does not guarantee continuous gas delivery to the immediate vicinity of the black hole.  From a dynamical perspective, bar strength is itself the outcome of secular angular-momentum redistribution within the galaxy. In isolated barred-galaxy simulations, stronger and longer bars form when more angular momentum is emitted by the inner disk and absorbed by resonant material in the outer disk or spheroidal components; the same exchange also causes the bar pattern speed to decrease with time \citep{Athanassoula2003,Lopez2024,Semczuk2024}. Therefore, a stronger large-scale bar signature should be interpreted primarily as evidence for efficient secular redistribution of angular momentum on galactic scales, not as a direct measure of the instantaneous gas inflow rate onto the black hole.

The final transport from circumnuclear scales to sub-100 pc scales requires additional angular momentum loss and can be affected by dynamical barriers such as inner Lindblad resonances, nuclear rings, central mass concentrations, or secondary non-axisymmetric structures \citep{Shlosman1989,Athanassoula1992,Combes2001,Englmaier2004,Kormendy2004, Athanassoula2013,Combes2019,Audibert2019}. In this picture, a bar may build up a circumnuclear gas reservoir without necessarily producing an enhancement of instantaneous black hole accretion.

Our results are consistent with this inner-transport bottleneck scenario. Barred NLS1s do not show enhanced optical AGN luminosity relative to unbarred
NLS1s, and the apparent bar strength, $e_{\max}$, is mildly anti-correlated with the optical continuum and broad Balmer-line luminosities within the
barred subsample. Since $e_{\max}$ is an apparent quantity affected by projection and host-galaxy structure \citep{Zou2014}, this trend should not be interpreted as direct evidence that intrinsically stronger bars suppress accretion. 

Moreover, bar ellipticity and bar length characterize different structural properties of a bar. The former primarily traces the apparent elongation and non-axisymmetric shape of the bar, whereas the latter characterizes its radial extent. Consistent with this distinction, we find no significant correlation between the AGN indicators and the normalized bar length, $a_{\rm bar}/{\rm FWHM}_{\rm eff}$, despite the mild anti-correlations found with $e_{\max}$. This difference further suggests that the $e_{\max}$--luminosity relation should not be interpreted as a simple dependence of AGN activity on the overall prominence or size of the bar.

Theoretical studies show that intrinsic bar strength depends on the history of angular momentum exchange between the disk, halo, and bulge, as well as on their mass distributions and velocity dispersions \citep{Athanassoula2003}. The observed $e_{\max}$--luminosity trend may therefore reflect a combination of bar evolutionary stage, host dynamics, and inner fueling efficiency.

The presence of the luminosity offset in the continuum, broad lines, and narrow lines may also provide a timescale clue. The continuum and broad lines
trace the more immediate ionizing output, whereas the NLR responds over longer timescales \citep[e.g.,][]{Peterson1993,Bennert2002}. The fact that the offset is also present in narrow H$\alpha$, narrow H$\beta$, and [\ion{O}{3}] suggests that the difference is not purely an instantaneous fluctuation confined to the innermost accretion flow. However, the present data cannot determine whether the offset reflects a long-lived secular difference or an episodic duty-cycle difference between barred and unbarred systems.

If the luminosity offset persisted over very long timescales, one might expect a corresponding difference in accumulated black hole growth. 
The absence of a significant $M_{\rm BH}$ difference suggests that the luminosity offset does not translate into a clearly detectable difference in
accumulated black hole growth within the uncertainties of single-epoch virial mass estimates \citep{Vestergaard2006,ShenKelly2010}. 
Larger samples, spatially resolved measurements of circumnuclear gas and star formation, and simulations that follow both bar evolution and black hole fueling will be needed to distinguish these
possibilities.

\subsection{Limitations}
We do not explicitly match the barred and unbarred subsamples in host stellar mass. Reliable stellar masses for Type~1 AGNs require dedicated AGN--host decomposition, and simple broad-band estimates can be affected by AGN continuum, emission-line contamination, and host-galaxy structure \citep[e.g.,][]{Kim2008}. In addition, stellar mass may be part of the physical connection between bars and host galaxies, rather than just a matching variable. Large-scale bars are connected to disk structure and secular evolution, so aggressively matching stellar mass could also remove physically relevant differences associated with bar formation. We therefore match the samples in redshift, inclination, image quality, and spatial resolution, which directly control bar detectability and observational selection.

A residual host-mass difference is unlikely to explain the direction of our main result. Since bar fractions generally increase with host stellar mass \citep{Skibba2012,Gavazzi2015}, any residual mass difference would most naturally make the barred subsample more massive on average. If the optical luminosity offset were driven mainly by host mass, for example through the $M_{\rm BH}$--$M_\star$ relation and comparable Eddington ratios, the barred subsample would then be expected to be the more luminous one. This is opposite to what we observe. Consistently, the two subsamples show statistically indistinguishable $M_{\rm BH}$ and Eddington-ratio distributions (Section~3.6). Thus, while the lack of direct host-mass matching remains a limitation, a residual host-mass difference is unlikely to account for the lower optical AGN luminosities of barred NLS1s.

Our bar classification relies on isophotal ellipse fitting, which becomes less sensitive when a bright nuclear point source dominates the inner isophotes. Since NLS1s are high-Eddington systems, some genuinely barred galaxies, especially the most AGN-dominated ones, may fail the bar criteria and be assigned to the unbarred subsample. This possible incompleteness would preferentially move AGN-bright barred systems into the unbarred subsample.
This effect would therefore tend to enhance the observed barred--unbarred difference. The observed luminosity offset should therefore be interpreted with
caution, as this potential bar incompleteness could create or enhance the difference between the two subsamples.

\section{Conclusions}

We have conducted a systematic comparison between barred and unbarred NLS1 galaxies using PS1-resolved matched subsamples controlled in redshift,
inclination, and image quality. Our main findings can be summarized as follows:

\begin{enumerate}

\item Barred NLS1s do not show enhanced optical AGN luminosity relative to their unbarred counterparts. Instead, the unbarred subsample is
modestly but consistently brighter in several optical AGN luminosity indicators, including $\lambda L_{\lambda}(5100~\mbox{\AA})$, broad
H$\alpha$, broad H$\beta$, narrow H$\alpha$, narrow H$\beta$, and [\ion{O}{3}]~$\lambda 5007$. The typical offsets are $\sim 0.1$--$0.25$ dex.

\item The luminosity offset is not accompanied by a significant difference in the main NLR ionization or density diagnostics. The
$\log([\mathrm{O\,III}]~\lambda 5007/\mathrm{H}\beta)$ ratio, O3N2 index, and [\ion{S}{2}]-based electron density are statistically consistent between
the barred and unbarred subsamples. The barred systems show a higher N2 ratio, but this is not accompanied by a significant enhancement of the
[\ion{N}{2}] luminosity itself.

\item The observed luminosity differences are unlikely to be driven primarily by differential dust attenuation, $\log M_{\rm BH}$, $\log\lambda_{\rm Edd}$,
radio-loud contamination, or X-ray obscuration. Both the narrow-line and broad-line Balmer decrements show no significant barred--unbarred difference.
The two subsamples also have statistically similar $\log M_{\rm BH}$, $\log\lambda_{\rm Edd}$, radio-loud fractions, and archival X-ray hardness
ratios.

\item Within the barred subsample, the apparent bar strength, quantified by $e_{\max}$, shows a mild anti-correlation with optical continuum and broad
Balmer-line luminosities. This trend is most clearly detected in systems with moderate inclination. No significant correlation is found between $e_{\max}$
and the narrow-line [\ion{O}{3}] luminosity.

\end{enumerate}

Taken together, these results suggest that large-scale stellar bars are not associated with enhanced instantaneous optical AGN luminosity in the matched NLS1 sample. A possible explanation is that bars can move gas into the central region of a galaxy, but this gas does not necessarily reach the black hole continuously. Gas may instead remain on circumnuclear scales before smaller-scale processes drive it further inward. In this case, barred systems can contain central gas reservoirs without showing stronger nuclear activity at the time of observation. The modestly lower optical luminosities of barred NLS1s therefore suggest that the bar--AGN connection is not a simple one. Large-scale stellar bars may affect where gas accumulates and when it is delivered inward, rather than directly increasing the instantaneous accretion power of the AGN.

\section*{Data Availability Statement}

The Pan-STARRS1 stack images used in this work were obtained through the STScI Pan-STARRS1 Image Cutout Service. Pan-STARRS1 catalog data
were obtained from the Pan-STARRS1 DR2 Catalog through the Mikulski Archive for Space Telescopes (MAST) at the Space Telescope Science Institute \citep{STScI2022PS1}. The matched barred and unbarred NLS1 samples underlying this analysis are provided with this article in a CSV format.

\section*{Acknowledgments}
P.-C.~Y. is supported by the Taiwan Astronomical Research Alliance (TARA) and Institute of Astronomy and Astrophysics, Academia Sinica (ASIAA), with funding from the National Science and Technology Council of Taiwan (NSTC 115--2740--M--008--001). TARA is committed to advancing astronomy in Taiwan and paving the way for the establishment of a national observatory. We gratefully acknowledge support from the AAS Publication Support Fund for the publication of this article.

The Pan-STARRS1 Surveys (PS1) and the PS1 public science archive have been made possible through contributions by the Institute for Astronomy, the University of Hawaii, the Pan-STARRS Project Office, the Max-Planck Society and its participating institutes, the Max Planck Institute for Astronomy, Heidelberg and the Max Planck Institute for Extraterrestrial Physics, Garching, The Johns Hopkins University, Durham University, the University of Edinburgh, the Queen's University Belfast, the Harvard-Smithsonian Center for Astrophysics, the Las Cumbres Observatory Global Telescope Network Incorporated, the National Central University of Taiwan, the Space Telescope Science Institute, the National Aeronautics and Space Administration under Grant No. NNX08AR22G issued through the Planetary Science Division of the NASA Science Mission Directorate, the National Science Foundation Grant No. AST--1238877, the University of Maryland, Eotvos Lorand University (ELTE), the Los Alamos National Laboratory, and the Gordon and Betty Moore Foundation. 

Funding for the Sloan Digital Sky Survey V has been provided by the Alfred P. Sloan Foundation, the Heising-Simons Foundation, the National Science Foundation, and the Participating Institutions. SDSS acknowledges support and resources from the Center for High-Performance Computing at the University of Utah. SDSS telescopes are located at Apache Point Observatory, funded by the Astrophysical Research Consortium and operated by New Mexico State University, and at Las Campanas Observatory, operated by the Carnegie Institution for Science. The SDSS web site is \url{https://www.sdss.org}.

SDSS is managed by the Astrophysical Research Consortium for the Participating Institutions of the SDSS Collaboration, including the Carnegie Institution for Science, Chilean National Time Allocation Committee (CNTAC) ratified researchers, Caltech, the Gotham Participation Group, Harvard University, Heidelberg University, The Flatiron Institute, The Johns Hopkins University, L'Ecole polytechnique f\'{e}d\'{e}rale de Lausanne (EPFL), Leibniz-Institut f\"{u}r Astrophysik Potsdam (AIP), Max-Planck-Institut f\"{u}r Astronomie (MPIA Heidelberg), Max-Planck-Institut f\"{u}r Extraterrestrische Physik (MPE), Nanjing University, National Astronomical Observatories of China (NAOC), New Mexico State University, The Ohio State University, Pennsylvania State University, Smithsonian Astrophysical Observatory, Space Telescope Science Institute (STScI), the Stellar Astrophysics Participation Group, Universidad Nacional Aut\'{o}noma de M\'{e}xico, University of Arizona, University of Colorado Boulder, University of Illinois at Urbana-Champaign, University of Toronto, University of Utah, University of Virginia, Yale University, and Yunnan University.  

This research has made use of the High Energy Astrophysics Science Archive Research Center (HEASARC), a service of the Astrophysics Science Division at NASA/GSFC, the Chandra Source Catalog, provided by the Chandra X-ray Center (CXC) as part of the Chandra X-ray Observatory mission, and the XMM-Newton 5XMM-DR15 Source Catalogue compiled by the institutes of the XMM-Newton Survey Science Centre selected by ESA.

The author acknowledges the use of AI-based tools during the preparation of this manuscript. ChatGPT (OpenAI) and Claude (Anthropic) assisted with language editing and the development of analysis code. All AI-assisted code was reviewed, tested, and verified by the author. These tools were not used independently to generate results, interpretations, or conclusions. All scientific content and conclusions presented in this work remain the sole responsibility of the author.

\vspace{5mm}

\software{
Python,
Astropy \citep{2013A&A...558A..33A,2018AJ....156..123A,2022ApJ...935..167A},
Astroquery \citep{Ginsburg2019},
IRAF \citep{Tody1986,Tody1993},
Matplotlib \citep{Hunter2007},
NumPy \citep{Harris2020},
pandas \citep{McKinney2010},
PyMC \citep{Salvatier2016},
PyNeb \citep{Luridiana2015},
SciPy \citep{Virtanen2020},
statsmodels \citep{Seabold2010},
scikit-learn \citep{pedregosa2011}
}

\bibliography{ms}{}
\bibliographystyle{aasjournal}

\end{document}